\documentclass[12pt]{article}
\usepackage[hmargin=2.cm,vmargin=3cm]{geometry}
 \usepackage{graphicx}
 \usepackage{amsmath}
 \usepackage{amssymb}
  \usepackage{enumerate}

\begin{document} \title{Measurements on relativistic quantum fields: I.  Probability assignment} \author {Charis Anastopoulos\footnote{anastop@physics.upatras.gr} and   Ntina Savvidou\footnote{ksavvidou@upatras.gr}\\
 {\small Department of Physics, University of Patras, 26500 Greece} }

\maketitle

\begin{abstract}
We present a new method for describing quantum measurements in relativistic systems that applies  
 (i)  to any QFT and for any field-detector coupling, (ii) to the measurement of any observable, and (iii) to arbitrary size, shape and motion of the detector.  We explicitly construct the probabilities associated to $n$ measurement events, while treating the spacetime coordinates of the events are random variables. These probabilities define  a linear functional of a $2n$ unequal time correlation function of the field, and thus, they are Poincar\'e covariant. The probability assignment  depends  on the properties of the measurement apparatuses,  their state of motion,  intrinsics dynamics,  initial states and  couplings to the measured field.   For each apparatus, this information is  contained in a function, the {\em detector kernel},  that   enters into the probability assignment. In a companion paper, we construct the detector kernel for different types of measurement.
\end{abstract}

\section{Introduction}

 We develop a general method for defining the probabilities associated to a sequence of measurements on a relativistic quantum field. The method applies to   any Quantum Field Theory (QFT) and for the measurement of any observable. Furthermore, it involves  a detailed mathematical modeling of the associated measurement apparatuses.

 We construct the probabilities corresponding to a sequence of measurements using the Quantum Temporal Probabilities (QTP) method \cite{AnSav12}.  QTP addresses   the time-of-arrival problem in quantum mechanics \cite{AnSav12, AnSav06}.
 It  inherits the rich
 temporal structure of   Histories theory, as it was developed by one of us (N.S.) \cite{Sav99}---implementing
 an important distinction   between the time parameter of Schr\"odinger equation and  the time variable characterizing a measurement event.  In the relativistic context \cite{Sav02,Sav04}, this  distinction is mirrored into one  between the parameters of spacetime translations and the spacetime coordinates associated to   a measurement record.
The QTP method  has also been applied for the temporal characterization of tunneling  \cite{AnSav08} and non-exponential decays \cite{An08}, and for calculating the response and correlations of particle detectors in non-inertial motion \cite{AnSav11}.

\bigskip

{\em Detectors in relativistic systems.}  Several important theorems about the general properties of measurements in relativistic systems have been proved by axiomatic   approaches to QFT  \cite{StrWi,Emch, Haag, Araki}. Nonetheless, there exist
  few concrete models of a fully relativistic treatment of the interactions
between a microscopic system and a measuring apparatus, and derivations of the associated probabilities. Almost all existing models treat the apparatus degrees of freedom using non-relativistic physics rather than QFT \cite{UnWa84, LPP92}---an  exception is Ref.  \cite{CoPi}.
This is the case for the most commonly employed  detector models, namely,  the Unruh-Dewitt \cite{Unruh, Dewitt} and the Glauber  \cite{Glauber}  detectors. The former  is mainly used for studying particle creation effects in moving frames. The latter defines the standard model of photo-detection theory, and it can be generalized in order to describe continuous photo-detection \cite{DaSi} and to address issues of relativistic causality \cite{MJF95}.

  In general, Poincar\'e covariant  unitary dynamics exists only if the interactions are expressed in terms of local quantum fields \cite{Weinb}. Models based on particle-field interactions may be very useful, but they have the potential of severely misrepresenting multi-partite systems when addressing issues of locality and causal propagation of information.

Note that in the present context, a "detector" or an "apparatus" is a elementary detecting element whose records can be correlated with a {\em single microscopic event}, for example,
   a single bubble in a bubble chamber or a wire segment in a wire chamber \cite{PeTe}.  Hence,  an experiment with $n$ distinct  records of observation,  requires $n$    distinct {\em independent} detecting elements, one for each record.   This is different from the common use
 of the word "detector"  (e.g., in high-energy experiments) that typically refers to a large collection of elementary detecting elements.

 %The novel features of our approach to relativistic quantum measurements are (i) the inclusion of the spacetime coordinates of any measurement event in the observed quantities and (ii) dual classical/quantum description of the detectors through a careful implementation of coarse-graining. In what follows, we proceed to an explanation of these points.

\bigskip

{\em Spacetime coordinates as observables.}
  In the axiomatic formulations of QFT,  observables are typically expressed in terms of
   operators  associated to bounded regions ${\cal O}$ in spacetime. These spacetime regions   are specified independently of the associated measurement outcomes. This means  that the spacetime coordinates of the  measurement outcomes (i.e., when and  where a record appeared in a detector) appear as external parameters in the associated probabilities and {\em not as   random variables}.

   There is little difference here  from non-relativistic quantum mechanics,   where time appears as a parameter of Schr\"odinger's equation and not as a random variable. To see this, one should recall that the probabilities provided by Born's rule are defined {\em at} an instant of time and not {\em with respect to} time. Also, in von Neumann's measurement theory, the time of measurement is predetermined by a `pulse' that switches on the interaction of the system with the measurement apparatus \cite{vNeu}.

 Nonetheless, the time of a measurement event {\em is} a random variable in many experimental set-ups. Consider, for example, a decay   $A \rightarrow B_1 + B_2 + \ldots$ of
an unstable particle $A$   into different product particles $B_i$. The time of decay is not directly observable, but the time at which one  of the product particles is recorded is observable. In general, the detection time is a random variable described by a probability density function that should be constructed according to the rules of quantum theory. Furthermore,  the product particles $B_i$ can be  detected at different spatial locations, because their directions of motion after the decay are random variables.  Hence, in this set-up
the spacetime coordinates of a detection event are genuine random variables.

The  long-standing debate about defining probabilities with respect to time  in quantum theory
  is best demonstrated by the time-of-arrival problem:
given an  initial wave function $|\psi_0 \rangle$ for a particle is centered around $x = 0$ and with a positive mean momentum, find the probability $P(t) \delta t$ that the particle is detected at distance $x = L$ at some moment between $t$ and $t+\delta t$. There is no canonical answer even to such an elementary question. Several different proposals exist and these agree only at the classical limit---for reviews see, Ref. \cite{ML, ToAbooks}.

%Another case of ambiguous  predictions of quantum theory for temporal variables is the specification of tunneling time, i.e., how long it  takes a particle to tunnel through a potential barrier \cite{tunnel1}.

 \bigskip

{\em Time in histories theory.}
 The novel feature of our approach to relativistic quantum measurements is the treatment of the spacetime coordinates $X$ of measurement events as random variables. This result follows from
  the
  key idea in Ref. \cite{Sav99} that `time' arises in quantum  theory in two natural ways: (i) as the parameter of causal ordering that distinguishes between past, present and future; and (ii) as an evolution parameter in dynamics. We present   a brief summary of this distinction between these two aspects of time.      Time translations are implemented by two distinct parameters, one of which refers to the kinematical set-up of the theory while the other refers solely to its dynamical behavior. These parameters correspond to two distinct types of time transformation. One refers to time as it appears in temporal logic; the other refers to time as it appears in the implementation of dynamical laws. For any specific physical system the action operator---a quantum analogue of the classical action functional---intertwines the two transformations.

 We emphasise the distinction between the notions of time evolution and that of logical time-ordering. The latter refers to the causal ordering of logical propositions about properties of the physical system. The corresponding parameter $t$ does not coincide with the notion of physical time---as, for example, is measured by a clock. Rather, it is an abstraction, which keeps only the ordering properties of physical time: i.e., it designates the sequence at which different events happen, as in the notion of time-ordered products in QFT.  When generalizing to relativistic systems, the logical time-ordering of events is expressed in terms of the coordinates $X$ of Minkowski spacetime, and  the two different transformation laws correspond to two different representations of the Poincar\'e group \cite{Sav02}. The analogous structures in classical general relativity are the spacetime diffeomorphism group and the Dirac algebra of constraints \cite{Sav04}.

 This conceptual distinction is essential for our definition of  probabilities for relativistic quantum measurements. In the usual description of QFT, spacetime points are viewed as parameters of the Poincar\'e group external to the system, and they cannot be treated as random variables. In contrast, there is no problem in treating the coordinates associated to logical time-ordering as random variables, indeed, the probability amplitudes are naturally {\em densities} with respect to the coordinates.

We must note here that, although originally these ideas about the notion of time were presented within  the Histories formalism, they were subsequently developed within different frameworks. The key aspects of this  theory  is presented in Ref. \cite{Sav10}. The QTP method is one example of application of these ideas outside a histories-based theory.

\bigskip

 {\em Coarse-grained description of the apparatus.}
Our approach to quantum measurements is  pragmatic rather than interpretational. We focus on a general mathematical formulation of relativistic quantum measurements, seeking to construct probabilities associated to measurement events in specific experimental set-ups. We do not propose or endorse a specific resolution of measurement problem and we make no commitment to a specific interpretation of quantum mechanics.
As in most operational descriptions of measurements, we employ   a dual quantum/classical description of the detectors. A novel feature of our method is the implementation of this duality through a careful treatment  of coarse-graining in the detector degrees of freedom.

    We describe a detector as a classical macroscopic system by specifying its associated world-tube in Minkowski spacetime. We also describe the detector   quantum-mechanically: it is associated to a Hilbert space ${\cal K}$ and the pointer variables are expressed in terms of operators on ${\cal K}$. Relativistic causality implies that the interactions between the detector and the microscopic system are local, in the sense that   the interaction Hamiltonian is a local functional of associated quantum fields. The coexistence of the classical and the quantum description follows from the postulate that the spacetime coordinates of a measurement event correspond to macroscopic quasi-classical variables of the detector. Here, we use the word `quasi-classical' in the sense of the decoherent histories approach to quantum mechanics  \cite{Omn1, Omn2, Gri, GeHa}, denoting coarse-grained quantum variables that satisfy appropriate decoherence conditions, and they approximately satisfy  classical evolution equations  \cite{GeHa, hartlelo}.

The coarse-graining of the space-time coordinates is important for the relativistic consistency of our approach.
  We assume that each detector  is sufficiently small in size, so that it  can be described in terms of  a {\em single} proper time parameter. This is an approximation, because in relativity different parts of an extended object move along different trajectories, and thus, they are characterized by different proper time variables. For a sufficiently coarse time variable,  the uncertainty
    in  the time of a measurement event is much larger than any proper-time ambiguity  due to the finite size of the detector.

\bigskip

 {\em General form of the probabilities.}
We express all probabilities associated to relativistic quantum measurements in terms of Positive-Operator-Valued Measures (POVMs).
 For a single measurement event, we define
 $\hat{\Pi}(X, \mu)$, where  $\mu$ denotes any  recorded observable other than the spacetime coordinates $X$ of the event (for example, spin or momentum).
  For $n$ different measurement events (for example, in a multipartite system), we define positive operators
  \begin{eqnarray}
  \hat{\Pi}(X_1, \mu_1; X_2, \mu_2; \ldots; X_n, \mu_n), \nonumber
   \end{eqnarray}
   where $X_i$ is the spacetime point and $\mu_i$ the recorded observable associated to the $i$-th measurement event, for $i = 1, 2, \ldots, n$.

  We  construct the positive operators $\hat{\Pi}(X_1, \mu_1; X_2, \mu_2; \ldots; X_n, \mu_n)$ through a careful modeling of the interaction between the microscopic system and the measurement apparatus. The associated probability densities   turn out to be  linear functionals of a   $2n$-point correlation function of the QFT that describes the  microscopic system. These correlation functions   involve $n$ time-ordered and $n$ anti-time-ordered components. The  probabilities transform covariantly under a change of the reference frame, provided that the underlying  QFT is Poincar\'e covariant.

  The probabilities depend explicitly on the properties of a measurement apparatus,  its state of motion, its intrinsics dynamics, its initial state and its coupling to the measured field.   For any apparatus, all this information is  contained in a function $R$, the {\em detector kernel},  that   enters into the probability assignment. In a companion paper \cite{AnSav15b}, we present a general model for detectors and identify explicitly the key component needed for the specification of $R$. The probabilities are uniquely defined from the knowledge of the correlation function of the field and the detector kernels for all apparatuses in an experimental setup.

 Thus, we obtain a very general description of relativistic  measurements, going well beyond the Glauber and Unruh-Dewitt models. Our method applies  to any QFT and to any {\em local}  field-detector coupling,  for the measurement of any observable together with the spacetime coordinates of events, and for all possible states of  motion of the detector.

%We find that all information about a  measurement apparatus can be codified in the following mathematical objects: (i) an embedding function that describes its world tube in spacetime, (ii) a POVM that describes the measurement of variables other than the spacetime coordinates of an event, (iii) the spatial and temporal resolution of the spacetime coordinates, and (iv) a function of time that describes the persistence of measurement records. These mathematical objects appear explicitly in the predicted probabilities. In general, each choice  corresponds to a physically different apparatus. Of course, as in non-relativistic systems, there are simplifying choices that correspond to {\em ideal measurements}. These correspond to probability distributions that are largely insensitive to specific properties of the measurement apparatus.

%The present paper is meant to develop the general formalism for relativistic quantum measurements, which will be applied to concrete problems in future publications. Here, we present three specific models for a detector that are suitable for  (i) time-of-arrival measurements, (ii) photo-detection (obtaining   Glauber-type photodetection at the appropriate limit) and (iii) spin measurements. Thus, our approach acts as a unifying formalism for different models in different branches of physics, ranging from Glauber's theory of photo-detection in optics, to the standard $S$-matrix theory  and to moving detectors of the Unruh-Dewitt type.

\bigskip

In this paper, we set up the formalism for describing relativistic measurements, we present the main modeling assumptions and derive  the general expressions for the probabilities associated to such measurements. Detailed models of the detectors and applications are presented in   the follow-up paper \cite{AnSav15b}.
The structure of the present paper is the following. In Sec. 2, we present the  QTP method for constructing quantum probabilities in which time is a random variable. In Sec. 3, we construct explicitly the models for relativistic detectors and the probabilities associated to $n$ measurement events. In Sec. 4, we present the general formula for the probabilities in relativistic quantum measurements and we analyse its properties. In Sec. 5, we summarise and discuss our results.

\section{The  Quantum Temporal Probabilities method}
In this section, we  present  the QTP approach for defining probability densities with respect to time \cite{AnSav12}, and we generalize it for   $n$-time measurements in a  way that is compatible with relativity. Some of this  material in this section overlaps with the presentation of Refs. \cite{AnSav12, AnSav11}, but  the history analysis of alternatives, the results on $n$ time measurements and the discussion of the quantum Zeno effect are  novel.

\subsection{Amplitudes for measurement events}

\subsubsection{Single measurement event}

We consider a composite physical system that consists of a microscopic and a macroscopic component. The microscopic component is the quantum system to be measured and the macroscopic component is the measuring device.

We denote the Hilbert space associated to the composite system
 by
   ${\cal H}$. We describe a measurement event as a transition between two complementary subspaces of ${\cal H}$. Hence, we consider a splitting of
 ${\cal H}$  into two subspaces: ${\cal H} = {\cal
H}_+ \oplus {\cal H}_-$. The subspace ${\cal H}_+$ describes the accessible states of the system given that the event under consideration is realized. For example, if the event is a detection of a microscopic particle by  an  apparatus,   the subspace ${\cal H}_+$ corresponds to all states of the apparatus compatible with the macroscopic record of detection.  We
denote  the projection operator onto ${\cal H}_+$ as $\hat{P}$ and the projector onto ${\cal H}_-$ as $\hat{Q} := 1  - \hat{P}$.

We will construct the probability density with respect to time that is associated to the transition of the system from ${\cal H}_-$ to ${\cal H}_+$. We consider transitions that are correlated with the emergence of a macroscopic record of observation. Such transitions are {\em logically
 irreversible}. Once they occur, and a measurement outcome has been recorded,   further time evolution of the system does not affect our knowledge that they occurred. Thus, they define the temporal ordering of events for the studied system.

After the transition has occurred,  a pointer variable $\lambda$ of the measurement apparatus takes a definite value. Let
$\hat{\Pi}(\lambda)$ be  positive operators that correspond to the different values of $\lambda$.  For example,
when considering transitions associated with particle detection, the projectors $\hat{\Pi}(\lambda)$  may be correlated  to the position, or to the momentum of the microscopic particle. Since $\lambda$ has a value only under the assumption that a detection event has occurred, the alternatives
  $\hat{\Pi}(\lambda)$ span the subspace
  ${\cal H}_+$ and not the full Hilbert space ${\cal H}$. Hence,
  \begin{eqnarray}
  \sum_\lambda \hat{\Pi}(\lambda) =
\hat{P}.
\end{eqnarray}
\medskip

Next, we construct probability amplitudes associated to specific values of transition time.   In particular, we
consider
 the probability amplitude $| \psi; \lambda, [t_1, t_2] \rangle$ that, given an initial ($t=0$) state $|\psi_0\rangle \in {\cal H}_-$, a transition occurs during the time interval $[t_1, t_2]$ and a value $\lambda$ for the pointer variable is obtained
 for
 some observable.

 We first consider the case of vanishingly small time
interval, i.e., we set $t_1 = t$ and $t_2 = t + \delta t$, and we  keep only leading-order terms with respect to $\delta t$. At times prior to $t$, the state
lies in ${\cal H}_-$. This is taken into account by evolving the initial state $|\psi_0 \rangle$ with the restricted propagator in ${\cal H}_-$,
 \begin{eqnarray}
 \hat{S}_t =  \lim_{N
\rightarrow \infty} (\hat{Q}e^{-i\hat{H} t/N} \hat{Q})^N, \label{restricted}
\end{eqnarray}
where $\hat{H}$ is the Hamiltonian of the composite system.

By assumption, the transition occurs at some instant within the time interval $[t, t+\delta t]$. Hence, there is no constraint in the propagation from $t$ to $t + \delta t$: propagation  is implemented by  the unrestricted evolution operator   $e^{-i \hat{H} \delta t} \simeq 1 - i \delta t \hat{H}$. At time $t + \delta t$, the event corresponding to $\hat{\Pi}(\lambda)$ is recorded, so the
amplitude is transformed by the action of $\sqrt{\hat{\Pi}({\lambda})}$. For times greater than $t + \delta t$, there is no constraint, so the amplitude
evolves as $e^{-i \hat{H} (T-t)}$  until some final moment $T$.

At the limit of small $\delta t$, the successive operations above yield
\begin{eqnarray}
|\psi_0; \lambda, [t, t+ \delta t] \rangle =  - i \, \delta t \, \,e^{-i\hat{H}(T - t)} \sqrt{\hat{\Pi}}(\lambda) \hat{H} \hat{S}_t |\psi_0
\rangle. \label{amp1}
\end{eqnarray}

 The amplitude $|\psi_0; \lambda, [t, t + \delta t] \rangle$ is proportional to $\delta t$. Therefore, it defines  a {\em density} with respect to time: $|\psi_0;  \lambda, t \rangle := \lim_{\delta t \rightarrow 0}
\frac{1}{\delta t} | \psi_0; \lambda, [t, t + \delta t] \rangle$. This is a key point in our analysis, because it  leads to the definition genuine probability densities with respect to time.

We also note that $t$ in Eq. (\ref{amp1}) refers to the instant that the transition took place. This does not coincide with the moment of time at which the amplitude is evaluated, namely, $T$. Thus, Eq. (\ref{amp1}) includes two variables for time. One variable labels the occurrence of an event and the other corresponds to Schr\"odinger's time evolution. This is in full accordance with the  proposal of Ref. \cite{Sav99} about time in quantum theory.

The time $t$ in Eq. (\ref{amp1}) is a quantum version of the "first-crossing time" or "first-passage time" that is defined in the theory of stochastic processes \cite{firstpass}. In quantum theory, first-passage times have been employed in path integrals and in relation to the time-of-arrival problem \cite{firstpass2}.

In histories theory, the amplitude (\ref{amp1}) is associated to an exhaustive and exclusive set of  alternatives \cite{AnSav06}. To see this, we first recall that a history is a  sequence of  properties about a system, defined at different instants of time. A history is represented mathematically by time-ordered sequence of projection operators. It is convenient to consider discrete-time histories, so we split the interval $[0, T]$ into $N$ time steps. A general  $N$-time history $\alpha$ is a string of projectors $(\hat{E}_1, \hat{E}_2, \ldots, \hat{E}_n)$, the index $i = 1, \ldots, N$ referring to the time step. Then, the following histories form an {\em exhaustive and exclusive} set of alternatives.
\begin{eqnarray}
\alpha_1 := (\hat{P}, \hat{1}, \hat{1}, \ldots, \hat{1}) , \alpha_2 :=  (\hat{Q}, \hat{P}, \hat{1} \ldots, \hat{1}), \alpha_3 := (\hat{Q}, \hat{Q}, \hat{P}, \ldots, \hat{1}), \ldots,  \nonumber \\
  \alpha_N :=  (\hat{Q}, \hat{Q}, \hat{Q}, \ldots, \hat{P}), \alpha_{\emptyset} :=   (\hat{Q}, \hat{Q}, \hat{Q}, \ldots, \hat{Q}). \label{a12n}
\end{eqnarray}
 The histories $\alpha_i$ correspond to a transition at the $i$-th time step, while the history $\alpha_{\emptyset}$ corresponds to no transition at all.  Each history $\alpha_i$ can be subdivided into alternatives corresponding to different values of $\lambda$. At the continuous limit, the  histories $\alpha_i$ correspond to the amplitude (\ref{amp1}). As shown in Ref. \cite{AnSav06}, the mutual exclusion of the histories $\alpha_i$ is crucial for (\ref{amp1}) defining a density with respect to time.

\medskip

We rewrite Eq. (\ref{amp1}) as

\begin{eqnarray}
 |\psi_0;  \lambda, t \rangle = - i
e^{- i \hat{H} T} \hat{C}(\lambda, t) |\psi_0 \rangle, \label{amp2}
\end{eqnarray}
where   the {\em class operator} $\hat{C}(\lambda, t)$ is
 \begin{eqnarray}
  \hat{C}(\lambda, t) := e^{i \hat{H}t} \sqrt{\hat{\Pi}}(\lambda)
\hat{H} \hat{S}_t. \label{class}
\end{eqnarray}

 Since the amplitude $|\psi_0;  \lambda, t \rangle $ is a density
with respect to   $t$,  integration over $t$ is well defined. This integration corresponds to the logical conjunction of several different histories $\alpha_i$ of Eq. (\ref{a12n}) \cite{hartlelo,Ish94}. Thus, the total amplitude that the transition occurred at {\em some moment} within a time interval $[t_1, t_2]$ is

\begin{eqnarray}
| \psi; \lambda, [t_1, t_2] \rangle = - i e^{- i \hat{H}T} \int_{t_1}^{t_2} d t \hat{C}(\lambda, t) |\psi_0 \rangle. \label{ampl5}
\end{eqnarray}

Eq. (\ref{ampl5}) involves the restricted propagator Eq. (\ref{restricted}) which may be difficult to compute in practice. However, there is a simplifying approximation. We note that
 for $[\hat{P}, \hat{H}] = 0$,  the amplitude  $|\psi_0; \lambda, t \rangle$ vanishes. We consider a Hamiltonian   $\hat{H} = \hat{H_0} + \hat{H_I}$, where
$[\hat{H}_0, \hat{P}] = 0$, and $H_I$ a perturbing interaction. To leading order in the perturbation,

\begin{eqnarray}
 \hat{C}(\lambda, t) = e^{i \hat{H}_0t} \sqrt{\hat{\Pi}}(\lambda) \hat{H}_I e^{-i \hat{H}_0t},
\label{perturbed}
\end{eqnarray}
and the restricted propagator $\hat{S}_t$ does not appear in the amplitude Eq. (\ref{ampl5}). In fact, all models for relativistic measurements we consider in this paper follow from the approximate amplitude (\ref{perturbed}).

\subsubsection{Multiple events}

Next, we generalize the procedure described in Sec. 2.1.1  for  a sequence of $n$ measurement events.

We represent a sequence of $n$ events  by a sequence of nested closed linear subspaces of the Hilbert space ${\cal H}$,
\begin{eqnarray} {\cal H}_0 \subset {\cal H}_1
\subset {\cal H}_2 \subset \hdots \subset {\cal H}_n ={\cal H}. \label{ordersub} \end{eqnarray}

 The $i$-th event corresponds to a transition from the subspace space ${\cal H}_{i-1}$ to the subspace
\begin{eqnarray}
{\cal V}_i = {\cal H}_{i}-
{\cal H}_{i-1}. \label{vi}
\end{eqnarray}
Hence, the Hilbert space ${\cal H}$ splits as a direct sum
\begin{eqnarray}
{\cal H} = {\cal H}_0 \oplus {\cal V}_1 \oplus \ldots \oplus {\cal V}_n.
\end{eqnarray}

To see this,  consider the system after the last event has taken place. Further time evolution  does not affect our knowledge of what has occurred, so there is no need to project the evolution into any subspace of ${\cal H}$. Suppose the last event corresponds to a transition into a subspace ${\cal V}_n$. Then prior to the $n$-th event, we must restrict time evolution into the subspace ${\cal H} - {\cal V}_n = {\cal H}_{n-1}$, because we know that the system was not in ${\cal V}_n$. Prior to the $(n-1)$-th event, we must restrict time evolution into ${\cal H}_{n-1} - {\cal V}_{n-1} = {\cal H}_{n-2}$, and so one until the full sequence (\ref{ordersub}) is recovered. 

%The construction of an exhaustive and exclusive set of histories for the $n$ events is a straightforward generalization of Eq. (\ref{a12n}).

As an aside remark, we note that  in Eq. (\ref{ordersub}) the temporal ordering of events  is represented in terms of the ordering by inclusion of closed linear subspaces  in the Hilbert space.  These two orderings are physically very different, the former refers to time the latter to coarse-graining operations, and in histories theory they are represented by different mathematical structures \cite{Ish94}.

\medskip

We denote the projectors corresponding to the subspaces  ${\cal H}_i$  by $\hat{Q}_i$, and  the projectors corresponding to ${\cal V}_i$
  by $\hat{P}_i = \hat{Q}_i - \hat{Q}_{i-1}$. By definition $[\hat{P}_i, \hat{P}_j] = 0$, i.e., the events are assumed to be independent. This is a natural assumption, because the  events correspond to readings of independent and uncorrelated measurement apparatuses.

  The $i$-th event is accompanied to macroscopic records $\lambda_i$ described by positive operators $\hat{\Pi}_i(\lambda_i)$ that satisfy
\begin{eqnarray}
\sum_{\lambda_i} \hat{\Pi}_i(\lambda_i) = \hat{P}_i.
\end{eqnarray}

Next, we compute the amplitude $|\psi_0; \lambda_1, [t_1, t_1 +\delta t_1]; \ldots ; \lambda_n, [t_n, t_n +\delta t_n] \rangle$ that is associated to the following proposition:  "The first event occurs during the time interval $[t_1, t_1 +\delta t_1]$ and an outcome $\lambda_1$ is recorded, the
second event  occurs during the time interval $[t_2, t_2+\delta t_2]$ and an outcome $\lambda_2$ is recorded, and so on, for all $n$ events." The instants of time are ordered:  $t_1 < t_2 < \ldots < t_n$.

The  reasoning that led  to Eq. (\ref{amp1}) also applies here for the construction of the amplitude. Taking a succession of a restricted propagation,   measurement and free propagation for each event, we obtain
\begin{eqnarray}
|\psi_0; \lambda_1, [t_1, t_1 +\delta t_1]; \ldots ; \lambda_n, [t_n, t_n
+\delta t_n] \rangle = \delta t_1 \delta t_2 \ldots \delta t_n |\psi_0; \lambda_1, t_1; \ldots
; \lambda_n, t_n \rangle ,
 \end{eqnarray}

 where the vector
  \begin{eqnarray}
  |\psi_0; \lambda_1, t_1; \ldots ; \lambda_n, t_n \rangle = (-i)^n e^{-i\hat{H}T} \hat{C}(\lambda_1, t_1; \lambda_2, t_2; \ldots ;
\lambda_n, t_n) |\psi_0\rangle, \label{nampl}
\end{eqnarray}
 is a  density with respect to each time variables $t_i$, and

 \begin{eqnarray}
  \hat{C}(\lambda_1, t_1; \lambda_2, t_2; \ldots ; \lambda_n, t_n) = e^{i\hat{H}t_n} \sqrt{\hat{\Pi}}_n(\lambda_n)
\hat{H}\hat{S}^{n-1}_{t_n-t_{n-1}} \ldots \nonumber \\
\sqrt{\hat{\Pi}}_2(\lambda_2) \hat{H}\hat{S}^1_{t_2-t_1} \sqrt{\hat{\Pi}}_1(\lambda_1) \hat{H}\hat{S}^{0}_{t_1}. \label{nclass}
 \end{eqnarray}
is the class operator for $n$ events.
The operators $\hat{S}^i_t$ in Eq. (\ref{nclass})
are the
restricted propagators   in  the subspaces ${\cal H}_i$, for $i = 1, 2, \ldots n-1$.

\subsubsection{Time-ordered amplitudes}

The amplitude Eq. (\ref{nampl}) is defined only for $t_1 < t_2 < \ldots < t_n$. The ordering of times corresponds to the ordering of the subspaces ${\cal H}_i$ by inclusion, Eq. (\ref{ordersub}). This means that when
integrating $t_1$ and $t_2$ over intervals $U_1$ and $U_2$, all points of $U_1$ must be prior to all points of $U_2$. This restriction reduces the domain of possible applications of the formalism, because it cannot describe systems in which the ordering of different events is not fixed a priori. To address this problem, we must add contributions from amplitudes that correspond to different subspace orderings. Thus, it is possible to define an amplitude for all possible values of
 $t_1, \ldots, t_n$.

 First, we consider the case of $n=2$.  Two events  correspond to a nested sequence  $ {\cal H}_0  \subset {\cal H}_1  \subset {\cal H}$. We   define the subspaces ${\cal V}_1 = {\cal H}_1
- {\cal H}_0$ and ${\cal V}_2 = {\cal H} - {\cal H}_1$. Let $\hat{Q}_0$ and $\hat{Q}_1$ be the projectors corresponding to ${\cal H}_{0}$ and  ${\cal H}_{1}$ respectively, and $\hat{P}_{1}$ and $\hat{P}_2$ the projectors corresponding to ${\cal V}_1$ and ${\cal V}_2$ respectively. By definition
\begin{eqnarray}
\hat{Q}_0 + \hat{P}_1 + \hat{P}_2 = \hat{1}.
\end{eqnarray}

The relevant class operator, Eq. (\ref{nclass}) is
\begin{eqnarray}
\hat{C}(\lambda, t; \lambda', t') =  e^{i\hat{H}t'} \sqrt{\hat{\Pi}}_2(\lambda') \hat{H}\hat{S}^1_{t'-t} \sqrt{\hat{\Pi}}_1(\lambda)
\hat{H} \hat{S}^0_{t}, \label{2class}
 \end{eqnarray} where the operators $\sqrt{\hat{\Pi}}_1(\lambda)$ and $\sqrt{\hat{\Pi}}_2(\lambda')$ are defined on ${\cal V}_1$ and ${\cal V}_2$, respectively.

The operator Eq. (\ref{2class}) is defined for $t < t'$. In order to extend its definition for $t > t'$, we  consider a situation where the record $\lambda'$ is prior in time to the record $\lambda$. This means that the
subspaces ${\cal V}_1$ and ${\cal V}_2$ are exchanged in the subspace ordering. To this end, we define ${\cal H}'_1 = {\cal H}_0 \oplus {\cal V}_2$, and consider the sequence ${\cal H}_0 \subset {\cal H}'_1 \subset {\cal
H}$. We will denote the projector associated to ${\cal H}'_1$ as $\hat{Q}'_1$.
The associated class operator is

\begin{eqnarray}
 \hat{C}'( \lambda, t; \lambda', t') =  e^{i\hat{H}t} \sqrt{\hat{\Pi}}_1(\lambda) \hat{H}\hat{S}^{\prime 1}_{t-t'} \sqrt{\hat{\Pi}}_2(\lambda') \hat{H} \hat{S}^{0}_{t'}.
\label{2classb }
 \end{eqnarray}

 Then, the time-ordered amplitude density
 \begin{eqnarray} |\psi_0; \lambda, t; \lambda', t'\rangle_T = (-i)^2 e^{i\hat{H}T} \hat{D}(\lambda, t; \lambda', t') |\psi_0\rangle \label{ampldd}
 \end{eqnarray}
 is  defined for all
values of $t$ and $t'$, where
\begin{eqnarray}
\hat{D}(\lambda, t; \lambda', t') = \theta(t'-t) \hat{C}(\lambda, t; \lambda', t')  + \theta(t-t') \hat{C}'( \lambda, t; \lambda', t'), \label{d12}
 \end{eqnarray}
and $\theta(t)$ is the usual step-function.

We clarify the logical structure of the  class operator (\ref{d12}) by writing its  associated set of histories. We split the time interval $[0, T]$ into $N$ steps and we define
\begin{eqnarray}
\alpha_{ij}&:=& \left(\hat{Q}_0, \hat{Q}_0, \ldots, \underbrace{\hat{P}_1}_{i\mbox{ \small -th step}}, \hat{Q}_1, \hat{Q}_1, \ldots,  \underbrace{\hat{P}_2}_{j\mbox{\small -th step}},  \hat{1}, \ldots \hat{1} \right) \nonumber\\
\alpha'_{ij} &:=& \left(\hat{Q}_0, \hat{Q}_0, \ldots, \underbrace{\hat{P}_2}_{i\mbox{ \small -th step}}, \hat{Q}'_1, \hat{Q}'_1, \ldots, \underbrace{\hat{P}_1}_{j\mbox{ \small -th step}},  \hat{1}, \ldots \hat{1} \right)\nonumber \\
\alpha_{i\emptyset} &:=& \left(\hat{Q}_0, \hat{Q}_0, \ldots, \underbrace{\hat{P}_1}_{i\mbox{ \small -th step}}, \hat{Q}_1, \hat{Q}_1, \ldots \hat{Q}_1\right)  \nonumber\\
\alpha'_{i\emptyset} &:=& \left(\hat{Q}_0, \hat{Q}_0, \ldots,  \underbrace{\hat{P}_2}_{i\mbox{ \small -th step}}, \hat{Q}'_1, \hat{Q}'_1, \ldots \hat{Q}'_1\right) \nonumber\\
\alpha_{\emptyset\emptyset}&:=& \left(\hat{Q}_0, \hat{Q}_0, \ldots \hat{Q}_0 \right).
\end{eqnarray}
The histories $\alpha_{ij}$ describe an event associated to $\lambda$ at the $i$-th step and an event associated to $\lambda'$ at the later time step $j$. The histories $\alpha'_{ij}$ describe an event associated to $\lambda'$ at the $i$-th step and an event associated to $\lambda$ at the later time step $j$. The histories $\alpha_{i\emptyset}$ and $\alpha'_{i\emptyset}$ describe an event associated to $\lambda$ and to $\lambda'$ respectively at the $i$-th time-step. The history $\alpha_{\emptyset\emptyset}$ describes no event. The set of histories above is exhaustive and exclusive. The  amplitude (\ref{ampldd}) corresponds to the logical conjunctions of histories $\alpha_{ij}$ and $\alpha'_{ij}$ at the continuum limit.

\medskip

Integration over $t$ and over $t'$ is well defined for the  amplitude (\ref{d12}). Thus, coarse-graining over time can be implemented also for  two measurement events.

The generalization to $n$ events is straightforward. The time-ordered amplitude for $n$ events
\begin{eqnarray}
|\psi_0; \lambda_1, t_1; \ldots ; \lambda_n, t_n \rangle_T = (-i)^n e^{-i\hat{H}T} \hat{D}(\lambda_1, t_1; \lambda_2, t_2; \ldots ;
\lambda_n, t_n) |\psi_0\rangle,
\end{eqnarray}
 is expressed in terms of class operator $\hat{D}(\lambda_1, t_1; \lambda_2, t_2; \ldots ; \lambda_n, t_n)$ that is obtained by time-ordering the operator (\ref{nclass}) for all possible orderings of the $n$ events.

  \subsubsection{Perturbative evaluation}

For a Hamiltonian of the form $\hat{H} = \hat{H}_0 + \hat{H}_I$ where $\hat{H}_0$ preserves the subspaces ${\cal H}_i$ and $\hat{H}_I$ is a small interaction, the class operator $\hat{C}$, Eq. (\ref{nclass})  simplifies,

\begin{eqnarray}
\hat{C}(\lambda_1, t_1; \lambda_2, t_2; \ldots ; \lambda_n, t_n) = e^{i \hat{H}_0 t_n} \sqrt{\hat{\Pi}}_n(\lambda_n)  \hat{H}_I e^{-i\hat{H}_0(t_n-t_{n-1})}\ldots
\nonumber \\
\times e^{-i\hat{H}_0(t_2-t_{1})}\sqrt{\hat{\Pi}}_n(\lambda_n) \hat{H}_I e^{-i\hat{H}_0 t_1}, \end{eqnarray}
Equivalently,
  \begin{eqnarray} \hat{C}(\lambda_1, t_1; \lambda_2, t_2; \ldots ; \lambda_n, t_n)  = \hat{A}_n(\lambda_n,
t_n) \ldots \hat{A}_2(\lambda_2, t_2) \hat{A}_1(\lambda_1, t_1)
\end{eqnarray}
is expressed in terms of the Heisenberg-picture operators
\begin{eqnarray}
\hat{A}_i(\lambda, t) = e^{i \hat{H}_0t} \sqrt{\hat{\Pi}}_i(\lambda)  \hat{H}_Ie^{-i
\hat{H}_0t}. \end{eqnarray}

 The time-ordered class operator is then
 \begin{eqnarray} \hat{D}(\lambda_1, t_1; \lambda_2, t_2; \ldots ; \lambda_n, t_n)  = T[\hat{A}_n(\lambda_n, t_n) \ldots \hat{A}_2(\lambda_2, t_2)
\hat{A}_1(\lambda_1, t_1)], \label{todn}
\end{eqnarray}
where $T[\ldots]$ denotes the standard  time-ordering operation for products of Heisenberg-picture operators.

\subsection{Temporal coarse-graining}

By Born's rule, the squared modulus of the amplitude  Eq. (\ref{amp2}) should define
the probability $P (\lambda, [t_1, t_2])\/$ that at some time in the interval $[t_1, t_2]$ a detection with outcome $\lambda$ occurred,
\begin{eqnarray}
 P(\lambda, [t_1, t_2])
:= \langle \psi; \lambda, [t_1, t_2] | \psi; \lambda, [t_1, t_2] \rangle =   \int_{t_1}^{t_2} \,  dt \, \int_{t_1}^{t_2} dt' Tr\left(\hat{C}(\lambda, t)
\hat{\rho_0}\hat{C}^{\dagger}(\lambda, t')\right)  , \label{prob1}
\end{eqnarray}
where $\hat{\rho}_0 = |\psi_0\rangle \langle \psi_0|$.

However, the quantities $P(\lambda, [t_1, t_2])$ do not define a probability measure with respect to time $t$, because they do not satisfy the Kolmogorov additivity axiom of probabilities. To see this, consider the
probability corresponding to an
 interval $[t_1, t_3] = [t_1, t_2] \cup [t_2,
t_3]$,
 \begin{eqnarray}
 P(\lambda, [t_1, t_3]) = P(\lambda, [t_1, t_2]) + P(\lambda, [t_2, t_3]) + 2 Re \left[ \int_{t_1}^{t_2} \,  dt \, \int_{t_2}^{t_3} dt' Tr\left(\hat{C}(\lambda, t)
\hat{\rho_0}\hat{C}^{\dagger}(\lambda, t')\right)\right]. \label{add}
\end{eqnarray}

The Kolmogorov additivity condition $P(\lambda, [t_1, t_3]) = P(\lambda, [t_1, t_2]) + P(\lambda, [t_2, t_3])$ fails, unless

\begin{eqnarray}
 2 Re \left[ \int_{t_1}^{t_2} \,  dt \, \int_{t_2}^{t_3} dt' Tr\left(\hat{C}(\lambda, t) \hat{\rho_0}\hat{C}^{\dagger}(\lambda, t')\right)\right] = 0 \label{decond}
 \end{eqnarray}
 In the consistent/decoherent histories framework, Eq. (\ref{decond}) is referred to as the
 {\em consistency condition} \cite{Omn1, Omn2, Gri}. It is the minimal condition necessary for defining a consistent probability measure for histories.

   Eq. (\ref{decond}) does not hold for generic choices of $t_1, t_2$ and $t_3$. However, we expect that it holds given a
   sufficient degree of coarse-graining. That is, we assume that there exists a   time-scale $\sigma$, such that the non-additive terms in Eq. (\ref{add}) are strongly
   suppressed if   $ |t_2 - t_1| >> \sigma$ and $|t_3 - t_2| >> \sigma$. This is a natural assumption  for a system that involves a macroscopic component such as a measuring apparatus  \cite{GeHa, hartlelo}. Then, Eq. (\ref{prob1}) defines a probability measure when restricted to intervals of size  larger than $\sigma$. The scale  $\sigma$ defines the absolutely minimal resolution of a measuring apparatus that is allowed by the rules of quantum mechanics.
     Estimates of $\sigma$ for specific detector models are given in Sec. 3 of the follow-up paper \cite{AnSav15b}.

   The probabilities with respect to  $\lambda$ are consistently defined, if
   \begin{eqnarray}
   \langle \psi; \lambda, [t_1, t_2] | \psi; \lambda', [t_1, t_2] \rangle \simeq  \delta_{\lambda \lambda'} \langle \psi; \lambda, [t_1, t_2] | \psi; \lambda, [t_1, t_2] \rangle, \label{decla}
   \end{eqnarray}
for $|t_2 - t_1| >> \sigma$. This is a constraint on the positive operators $\hat{\Pi}(\lambda)$ that can represent a record $\lambda$. In general, Eq. (\ref{decla}) implies a restriction to  highly coarse-grained observables $\hat{\Pi}(\lambda)$. This is to be expected since $\lambda$ refers to a {\em macroscopically distinguishable record} on a measurement apparatus. Equivalently, Eq. (\ref{decla}) can be viewed as a {\em definition} of what it means for a class of positive operators $\hat{\Pi}(\lambda)$ to represent a macroscopic record of observation.

\subsection{Probabilities for measurement events}

\subsubsection{Single event}

We   define the time-of-transition probabilities by smearing the amplitudes
 Eq. (\ref{amp2}) with respect to the coarse-graining  time-scale  $\sigma$ rather than using  sharp time-intervals, as in Eq. (\ref{prob1}). Then, the time-of-transition  probabilities are expressed in terms of
 densities
of a continuous time variable.

To this end, we introduce a family of functions $f_{\sigma}(s)$,  localized around $s = 0$ with width $\sigma$, and normalized so that $\lim_{\sigma \rightarrow 0} f_{\sigma}(s) = \delta(s)$. It is convenient to employ  the
Gaussians
\begin{eqnarray} f_{\sigma}(s) = \frac{1}{\sqrt{2 \pi \sigma^2}} e^{-\frac{s^2}{2\sigma^2}}, \label{gauss} \end{eqnarray}
 even though any family of approximate delta functions is adequate.

  We must keep in mind that the functions $f_{\sigma}$ must approximate the delta function of the time interval $[0, T]$.
 This means that they should satisfy $f_{\sigma}(0) = f_{\sigma}(T) = 0$. The Gaussians are good approximate delta functions only if   $\sigma/T << 1$.

 The Gaussians Eq. (\ref{gauss}) satisfy the following equality.

\begin{eqnarray}
\sqrt{f_{\sigma}(t-s) f_{\sigma}(t-s')} = f_{\sigma}(t - \frac{s+s'}{2}) g_{\sigma}(s-s'), \label{eq2}
\end{eqnarray}

where

\begin{eqnarray}
g_{\sigma}(s) = e^{-\frac{s^2}{8\sigma^2}}. \label{gsig}
 \end{eqnarray}

Using the functions $f_{\sigma}$, we  define the smeared amplitude $|\psi_0; \lambda, t\rangle_{\sigma}$ that is localized
 around the time $t$, as
\begin{eqnarray}
|\psi_0; \lambda, t\rangle_{\sigma} := \int ds \sqrt{f_{\sigma}(s -t)} |\psi_0; \lambda, s \rangle =  \int ds \sqrt{f_{\sigma}(s - t)} \hat{C}(\lambda, s) |\psi_0 \rangle, \label{smearing}
\end{eqnarray}

 The modulus-
squared amplitudes
\begin{eqnarray}
 \bar{P}(\lambda, t) = {}_{\sigma}\langle \psi_0; \lambda, t|\psi_0; \lambda, t\rangle_{\sigma} = \int ds ds' \sqrt{f_{\sigma}(s-t) f_{\sigma}(s'-t)} Tr \left[\hat{C}(\lambda, s)
 \hat{\rho}_0 \hat{C}^{\dagger}(\lambda, s')\right] \label{ampl6}
\end{eqnarray}
 define a probability measure: they are of the form
 $Tr[\hat{\rho}_0 \hat{\Pi}(\lambda, t)]$,
where
\begin{eqnarray}
\hat{\Pi}(\lambda, t) = \int ds ds' \sqrt{f_{\sigma}(s-t) f_{\sigma}(s'-t)} \hat{C}^{\dagger}(\lambda, s') \hat{C}(\lambda, s) \label{povm2}
\end{eqnarray}
is a density with respect to both variables
$\lambda$ and $t$.

Using Eq. (\ref{eq2}), and setting $t' = (s+s')/2$, $\tau = s -s'$, Eq. (\ref{ampl6}) becomes
\begin{eqnarray} \bar{P}( \lambda, t) = \int dt' f_{\sigma}(t-t') P(\lambda, t'), \label{smearing2}
\end{eqnarray}
 where
\begin{eqnarray} P(\lambda, t) =\int d \tau g_{\sigma}(\tau) \left[\hat{C}(\lambda, t+\frac{\tau}{2})
 \hat{\rho}_0 \hat{C}^{\dagger}(\lambda, t- \frac{\tau}{2})\right] \label{pp}
\end{eqnarray}

The probability distribution $\bar{P}_{\sigma}$   is obtained by coarse-graining through convolution the classical probability distribution $P$ at a scale of $\sigma$.  For systems monitored at a
time-scale much larger than $\sigma$ the two probability distributions essentially coincide. In that case, the probability density $P$ may be employed instead of $\bar{P}$.

Moreover, if the resolution scale $\sigma$ is much
larger than any timescale characterizing the microscopic system, we can take the limit $\sigma \rightarrow \infty$ in Eq. (\ref{pp}), by setting $g_{\sigma} = 1$. The resulting probability distribution \begin{eqnarray}
P(\lambda, t) = \int d\tau  Tr\left[\hat{C}(\lambda, t+\frac{\tau}{2})
 \hat{\rho}_0 \hat{C}^{\dagger}(\lambda, t- \frac{\tau}{2})\right] \label{pp2}
\end{eqnarray} is independent of the coarse-graining scale $\sigma$.

\subsubsection{The event of no detection and the quantum Zeno effect}

The operator $\sum_{\lambda} \int_0^{\infty} dt  \hat{\Pi} (\lambda, t)$ corresponds to the total probability that an event has been recorded in the time interval $[0, \infty)$. Consequently, the operator
\begin{eqnarray}
\hat{\Pi} _{\emptyset} = \hat{1} -  \sum_{\lambda} \int_0^{\infty} dt\lambda \hat{\Pi} (\lambda, t), \label{nodet}
\end{eqnarray}
 corresponds to the alternative $ \emptyset$ that no detection took place. The lack of a measurement record may be due to the fact that some of the particles in the statistical ensemble "missed" the detector, or it may be due to a  non-zero probability that the interaction of the microscopic  particles with the apparatus leaves no record.

 If the operator $\hat{\Pi} _{\emptyset} $ is positive, then $\hat{\Pi} _{\emptyset} $ together with the positive  operators Eq. (\ref{povm2})
 define a   POVM that is associated to a complete set of alternatives. However, the positivity of $\hat{\Pi} _{\emptyset} $ cannot be guaranteed by the assumptions that have been made so far. The problem arises from the properties of   the restricted propagator $\hat{S}_t$ of Eq. (\ref{restricted}), and it is related with the quantum Zeno effect \cite{MiSu77}.

To avoid inessential complications, we  consider a set-up where only the event of detection is recorded and no variable $\lambda$ appears. Assuming that $t \in [0, T]$, the associated probability density is

\begin{eqnarray}
 \bar{P}(t) =   \int_0^T ds \int_0^Tds' \sqrt{f_{\sigma}(s-t) f_{\sigma}(s'-t)} Tr \left[\hat{C}( s)
 \hat{\rho}_0 \hat{C}^{\dagger}( s')\right],
\end{eqnarray}
where $\hat{C}(s) =  e^{i \hat{H}s} \hat{P}
\hat{H} \hat{S}_s$. The functions $f_{\sigma}$ are approximate delta functions in the interval $[0, T]$.

We find that
\begin{eqnarray}
 \int_0^T dt \bar{P}(t) \leq  \int_0^T dt \left(\sup_{s,s'} \sqrt{f_{\sigma}(s-t) f_{\sigma}(s'-t)} \right) \int_0^T ds \int_0^T ds' Tr \left[\hat{C}( s)
 \hat{\rho}_0 \hat{C}^{\dagger}( s')\right]. \label{pbound}
\end{eqnarray}
The maximum of $\sqrt{f_{\sigma}(s-t) f_{\sigma}(s'-t)}$ is achieved for $s = s'$. Hence, the integral over $t$ in Eq. (\ref{pbound}) is $\int_0^T ds f_{\sigma}(s) = 1$, and we obtain
\begin{eqnarray}
 \int_0^T dt \bar{P}(t) \leq Tr \left[\hat{C}_+
 \hat{\rho}_0 \hat{C}^{\dagger}_+\right],
\end{eqnarray}
where $\hat{C}_+ = \int_0^T ds \hat{C}(s)$. Hence, if $Tr \left[\hat{C}_+
 \hat{\rho}_0 \hat{C}^{\dagger}_+\right] \leq 1$ for all $\hat{\rho}_0$, then $Tr(\hat{\rho}_0 \hat{\Pi}_{\emptyset})\geq 0$, i.e., the operator $\hat{\Pi}_{\emptyset}$ is positive.

However, the condition $Tr \left[\hat{C}_+
 \hat{\rho}_0 \hat{C}^{\dagger}_+\right] \leq 1$ is not guaranteed. This is best seen using arguments from the decoherent histories approach. The operator $\hat{C}_+$ represents the history  $\alpha_+$ that a detection event takes place within the time interval $[0, T]$; $\alpha_+$ is the logical disjunction of all histories $\alpha_i$ on Eq. (\ref{a12n}). The negation of $\alpha_+$ is the history $\alpha_{\emptyset}$ that no detection event took place during $[0, T]$ and it is represented by the class operator
$\hat{C}_{\emptyset} = \hat{S}_T$. Since the histories $\alpha_+$  are exhaustive and exclusive, for any initial state $\hat{\rho}_0$ the following identity holds
\begin{eqnarray}
Tr \left( \hat{C}_+^{\dagger} \hat{\rho}_0 \hat{C}_+\right) + Tr \left( \hat{C}_{\emptyset}^{\dagger} \hat{\rho}_0 \hat{C}_{\emptyset}\right) + 2 \mbox{Re} Tr \left( \hat{C}_1^{\dagger} \hat{\rho}_0 \hat{C}_{\emptyset}\right) = 1. \label{normhis}
\end{eqnarray}
The identity (\ref{normhis}) is  the normalization condition for the decoherence functional \cite{Ish94}.

 The problem is that the restricted propagator is unitary in the subspace $\hat{H}_-$, i.e., it satisfies $\hat{S}_t \hat{S}_t^{\dagger} = \hat{Q}$ \cite{MiSu77}.
 Hence, for any initial state $\hat{\rho}_0$ with support only on ${\cal H}_0$,
 \begin{eqnarray}
 Tr \left( \hat{C}_{\emptyset}^{\dagger} \hat{\rho}_0 \hat{C}_{\emptyset}\right) = Tr \left( \hat{S}_T^{\dagger} \hat{\rho}_0 \hat{S}_T\right) = 1.
 \end{eqnarray}
  Eq. (\ref{normhis}) implies that
 \begin{eqnarray}
 Tr \left( \hat{C}_+^{\dagger} \hat{\rho}_0 \hat{C}_+\right)  = 2 \mbox{Re} Tr \left( \hat{C}_+^{\dagger} \hat{\rho}_0 \hat{C}_{\emptyset}\right). \label{condaa}
 \end{eqnarray}
The r.h.s. of Eq. (\ref{condaa}) is, in general, bounded above by $2$, so that there is no guarantee that  $Tr \left( \hat{C}_+^{\dagger} \hat{\rho}_0 \hat{C}_+\right) \leq 1$, so that $\hat{\Pi} _{\emptyset} $ is positive.

In the consistent histories approach, the probabilities are well defined only if
\begin{eqnarray}
\mbox{Re} Tr \left( \hat{C}_+^{\dagger} \hat{\rho}_0 \hat{C}_{\emptyset}\right) = 0,
 \end{eqnarray}
 which means that  $Tr \left( \hat{C}_+^{\dagger} \hat{\rho}_0 \hat{C}_+\right) = 0$, i.e., the particle is never detected. This may be viewed as a fault of the consistent histories approach. However, in Ref. \cite{HalYe} it was shown that there may be alternative definitions of the restricted propagator that involve a regularization time scale so that the operator $\hat{S}_T$ is not unitary  in $\hat{H}_-$. Then,   $Tr \left( \hat{C}_{\emptyset}^{\dagger} \hat{\rho}_0 \hat{C}_{\emptyset}\right) $ may be appreciably different from zero in the physically relevant regime.

 The results of \cite{HalYe} suggest that we could regularize the restricted propagator $\hat{S}_t$, perhaps by making it dependent on the temporal coarse-graining parameter $\sigma$---see, also Ref. \cite{AlMu12}. It is plausible that for  a suitably regularized expression $\hat{S}^{\sigma}_t$, we would obtain
\begin{eqnarray}
\mbox{Re} Tr \left( \hat{C}_+^{\dagger} \hat{\rho}_0 \hat{S}_T^{\sigma}\right) \simeq 0. \label{conditionsuf}
\end{eqnarray}
Then by Eq. (\ref{normhis}), we obtain $Tr \left( \hat{C}_+^{\dagger} \hat{\rho}_0 \hat{C}_+\right) + Tr \left( \hat{C}_{\emptyset}^{\dagger} \hat{\rho}_0 \hat{C}_{\emptyset}\right) = 1$. Hence, $Tr \left( \hat{C}_+^{\dagger} \hat{\rho}_0 \hat{C}_+\right)  \leq 1$ and the positivity of $\hat{\Pi}_{\emptyset}$ is guaranteed.

The construction of a regularized restricted propagator so that Eq. (\ref{conditionsuf}) holds is a {\em sufficient} condition for the definition of a POVM that includes the alternative $\hat{\Pi}_{\emptyset}$. However, it is not a necessary condition. In some cases, the POVM is well defined even without such a regularization \cite{AnSav06}. Alternatively, we could employ arguments analogous to those of Ref. \cite{Sok13}, and
{\em postulate} that the inequality
\begin{eqnarray}
Tr \left( \hat{C}_+^{\dagger} \hat{\rho}_0 \hat{C}_+\right) \leq 1
 \end{eqnarray}
 should be satisfied in any physically consistent measurement scheme.

We emphasize that the issue here is not that negative probabilities appear in physically relevant systems. In  fact, they have not appeared in any system that we have studied with  this method so far. The issue is how   to {\em guarantee} that they will not appear in any conceivable application of the formalism.

In any case, problems due to  the quantum Zeno effect  do not appear in  the results of this paper and its follow-up.  Our models for relativistic quantum measurements that   rely on the perturbative evaluation of the probabilities through Eq. (\ref{perturbed}) in which the restricted propagator $\hat{S}_t$ does not appear. Thus, the details of constructing a regularized propagator $\hat{S}_t^{\sigma}$ that satisfies Eq. (\ref{conditionsuf}) do not affect any physical predictions at this level of approximation.

\subsubsection{Multiple events}

Next, we  derive the probability density $\bar{P}_{\sigma}(\lambda_1, t_1; \lambda_2, t_2; \lambda_n, t_n)$ for $n$ events at times $t_1, t_2,
\ldots, t_n$ and leaving records corresponding to $\lambda_1, \lambda_2, \ldots, \lambda_n$.
Using the reasoning that led to Eqs. (\ref{smearing2}) and (\ref{pp}),  we obtain

\begin{eqnarray} \bar{P}(\lambda_1, t_1; \lambda_2, t_2; \lambda_n, t_n) = \int ds_1 ds_2 \ldots ds_n f_{\sigma}(t_1-s_1)
f_{\sigma}(t_2-s_2) \ldots f_{\sigma}(t_n - s_n) \nonumber \\ \times P(\lambda_1,s_1; \lambda_2, s_2; \ldots ; \lambda_n, s_n), \end{eqnarray}
 where

\begin{eqnarray} P (\lambda_1,t_1; \lambda_2, t_2; \ldots ;
\lambda_n, t_n) = \int d \tau_1 d \tau_2 \ldots d\tau_n g_{\sigma}(\tau_1) g_{\sigma}(\tau_2) \ldots g_{\sigma}(\tau_n)
 \nonumber \\
Tr\left[\hat{D}(\lambda_1, t_1+\frac{\tau_1}{2}, \lambda_2, t_2+\frac{\tau_2}{2}, \ldots , \lambda_n, t_n+\frac{\tau_n}{2} )
 \hat{\rho}_0 \hat{D}^{\dagger}(\lambda_1, t_1-\frac{\tau_1}{2}, \lambda_2, t_2-\frac{\tau_2}{2}, \ldots , \lambda_n, t_n-\frac{\tau_n}{2} )\right] \label{npp}
\end{eqnarray}

The time-ordered class operator $\hat{D}(\lambda_1, t_1; \ldots ; \lambda_n, t_n)$ is defined in Sec. 2.2.

 If the resolution scale $\sigma$ is much larger than any
timescale characterizing the microscopic system, we can take the limit $\sigma \rightarrow \infty$ in Eq. (\ref{npp}). The resulting probability distribution \begin{eqnarray} P (\lambda_1,t_1; \lambda_2, t_2; \ldots ;
\lambda_n, t_n) = \int d \tau_1 d \tau_2 \ldots d\tau_n Tr\left[\hat{D}(\lambda_1, t_1+\frac{\tau_1}{2}, \lambda_2, t_2+\frac{\tau_2}{2}, \ldots , \lambda_n t_n+\frac{\tau_n}{2} ) \right. \nonumber\\ \left.
 \hat{\rho}_0 \hat{D}^{\dagger}(\lambda_1, t_1-\frac{\tau_1}{2}, \lambda_2, t_2-\frac{\tau_2}{2}, \ldots , \lambda_n, t_n-\frac{\tau_n}{2} )\right] \label{npp2}
\end{eqnarray} is independent of the coarse-graining time-scale $\sigma$.

Such $n$-time measurements are represented by positive operators
\begin{eqnarray}
\hat{\Pi}_n (\lambda_1,t_1; \lambda_2, t_2; \ldots ;
\lambda_n, t_n) =  \int d \tau_1 d \tau_2 \ldots d\tau_n g_{\sigma}(\tau_1) g_{\sigma}(\tau_2) \ldots g_{\sigma}(\tau_n)\nonumber \\
\hat{D}^{\dagger}(\lambda_1, t_1+\frac{\tau_1}{2}, \lambda_2, t_2+\frac{\tau_2}{2}, \ldots , \lambda_1, t_n+\frac{\tau_n}{2} ) \hat{D}(\lambda_1, t_1+\frac{\tau_1}{2}, \lambda_2, t_2+\frac{\tau_2}{2}, \ldots , \lambda_1, t_n+\frac{\tau_n}{2} ).
\end{eqnarray}

We also write the operators $\hat{\Pi}_n (\lambda_1,t_1; \lambda_2, t_2; \ldots ; \emptyset_i ; \ldots ;
\lambda_n, t_n)$  corresponding to the $i$-th event not occurring within the time interval $[0, T]$ while all other events have occurred and given definite results
\begin{eqnarray}
\hat{\Pi}_n (\lambda_1,t_1; \lambda_2, t_2; \ldots ; \emptyset_i ; \ldots ;
\lambda_n, t_n)  = \hat{\Pi}_{n-1} (\lambda_1,t_1; \lambda_2, t_2; \ldots ;
\lambda_n, t_n)
\nonumber \\
- \sum_{\lambda_i} \int_0^T dt_i \hat{\Pi}_n (\lambda_1,t_1; \lambda_2, t_2; \ldots ; \lambda_i, t_i ; \ldots ;
\lambda_n, t_n).
\end{eqnarray}
As in the definition of the operator $\hat{\Pi}_{\emptyset}$ for a single event, special care must be taken in the definitions above, possibly involving a regularization of the restricted propagators. Again, such details are not necessary when employing the perturbative expressions (\ref{todn}) for the operators $\hat{D}(\lambda_1, t_1; \lambda_2, t_2; \ldots ; \lambda_n, t_n)$. Similarly, we define positive operators that correspond to  two events not occurring during $[0, T]$, and so on, until we define the operator $\hat{\Pi}_n(\emptyset; \emptyset; \ldots; \emptyset)$ that none of the $n$ events have occurred
\begin{eqnarray}
\hat{\Pi}_n(\emptyset; \emptyset; \ldots; \emptyset)  = \hat{1} - \sum_{\lambda_1, \ldots \lambda_n} \int_0^T dt_1 \ldots \int_0^T dt_n \hat{\Pi}_n (\lambda_1,t_1; \lambda_2, t_2; \ldots ;
\lambda_n, t_n).
\end{eqnarray}

Note that the operator $\hat{\Pi}_n(\emptyset; \emptyset; \ldots; \emptyset) $ for none of $n$ events happening is different from the operator $\hat{\Pi}_{n'}(\emptyset; \emptyset; \ldots; \emptyset)$ for none of $n'\neq n$ events happening. This is not paradoxical in an operational setting, where the events are associated to macroscopic records in an experimental set-up. An experiment set-up to record a maximum of $n$ measurement events involves  a different physical configuration from an experiment that is set-up to record a maximum of $n'$ measurement events.

\section{Modeling  relativistic quantum measurements}

Next, we employ the QTP method that was developed in Sec. 2  in order to construct a theory for relativistic quantum measurements. The system under consideration is a quantum field interacting with $n$ independent measuring apparatuses. The key point is that the events/transitions are defined solely with reference to the apparatuses' degrees of freedom. Each event corresponds to a macroscopic record that is left in a single apparatus. Hence, the measurement events define disjoint alternatives, which is the essential requirement for defining the relevant class operators---see, Sec. 2.1.2. We remind the reader that, as explained in the Introduction, by "detector" we refer to detecting elements correlated with single records of observation.

In what follows, we describe the main modeling assumptions, i.e., we identify the Hilbert spaces and operators that enter into the probability assignment of the QTP method.

\subsection{Subspaces associated to  $n$ measurement events}
We   consider an experimental set-up in which $n$ distinct and non-interacting apparatuses that record properties of a quantum field. We label the apparatuses by an index $i = 1,2, \ldots, n$. A Hilbert space ${\cal K}_i$ is associated to each apparatus. In the QTP method, a measurement event is associated to a transition between   two complementary subspaces. Hence, we assume that each Hilbert space
  ${\cal K}_i$  splits as
\begin{eqnarray}
{\cal K}_i = {\cal K}_i^- \oplus
{\cal K}_i^+. \label{spliti}
\end{eqnarray}

 In Eq. (\ref{spliti}), the subspace  ${\cal K}_i^-$   corresponds to the absence and  the subspace ${\cal K}_i^+$  to the presence of a macroscopic measurement record.  %For example, if the detector is identified with a silver halide crystal in a photographic film, the subspace ${\cal K}^+$ corresponds to all states compatible with the presence of a macroscopic agglomerate of neutral silver particles and ${\cal K}^-$ corresponds to all states
  We denote by $\hat{E}_i$ the projector associated to the subspace ${\cal K}_i^+$.  We assume that the initial state $|\omega_i\rangle$  of each  apparatus  lies in ${\cal K}_i^-$.

As an example of the subspaces above, we consider detection in a bubble chamber. Let ${\cal K}$  represent the Hilbert space associated to an element of the fluid, subject to suitable boundary conditions (pressure, temperature and so on). Let $\hat{R}$ be an operator that corresponds to the radius of the largest bubble in the fluid. Then,
  ${\cal K}_-$ is the subspace corresponding to all eigenvalues of $\hat{R}$ smaller than the critical radius that leads to the  amplification of the bubble and ${\cal K}_+$  corresponds to all eigenvalues of $\hat{R}$ larger than the critical radius.

%The self-dynamic of the apparatus is described by an one-parameter family of unitary operators $\hat{U}_{0i}(t)$. Due to the possible motion of the detectors, the evolution operators $\hat{U}_{0i}(t)$ may not be expressed in terms of a time-independent Hamiltonian of the apparatuses' degrees of freedom. We will refrain from giving an explicit form of $\hat{U}_{0i}(t)$ until Sec. 3.2, where we will present an explicit modeling of the measuring apparatus. At the moment, we only assume that $\hat{U}_{0i}(t)$ commutes with $\hat{E}_i$.

The measured  quantum system  is described by a quantum field theory defined on a Hilbert space ${\cal F}$. We denote the   Heisenberg-picture field operators as $\hat{\Phi}_r(X) := \hat{\Phi}_r({\bf x}, t)$, where
$r$ is a collective index that may include both spacetime and internal indices. The fields $\hat{\Phi}_r(X)$ may include both bosonic and fermionic components, and they may be either free or interacting.

The defining feature of a relativistic system is the existence of a unitary representation of the Poincar\'e group on
the Hilbert space ${\cal F}$. A unitary operator $\hat{U}(\Lambda, a)$ is associated to each element $(\Lambda, a)$  of the  Poincar\'e group, so that the fields $\hat{\Phi}_a(X)$ transform as

\begin{eqnarray}
\hat{\Phi}_r(X) \rightarrow \hat{U}^{\dagger}(\Lambda, a) \hat{\Phi}_r(X) \hat{U}(\Lambda, a) = D_{r}^{r'}(\Lambda) \hat{\Phi}_{r'}(\Lambda^{-1} X - a),  \label{poincare1}
\end{eqnarray}
for some matrix $D_r^{r'}(\Lambda)$.

%We denote the generator of time translations, the QFT Hamiltonian on ${\cal F}$, as $\hat{H}^{\Phi}$.

The Hilbert space  ${\cal H}$ describing the total system including the quantum fields and the measurement devices is

\begin{eqnarray}
 {\cal H} = {\cal F} \otimes {\cal K}_1 \otimes {\cal K}_2 \ldots \otimes {\cal K}_n. \label{totalH}
\end{eqnarray}

When the measurements event are ordered in such a way that the apparatus 1 detects first, the apparatus 2 detects second and so on, we define a nested sequence of Hilbert subspaces ${\cal H}_i$  of the type  Eq.  (\ref{ordersub}),
\begin{eqnarray}
{\cal F} \otimes {\cal K}_1^- \otimes {\cal K}_2^- \ldots \otimes {\cal K}_n^- \subset {\cal F} \otimes {\cal K}_1 \otimes {\cal K}_2^- \ldots \otimes {\cal K}_n^- \subset \ldots \nonumber \\ \subset {\cal F} \otimes {\cal
K}_1 \otimes {\cal K}_2 \ldots \otimes {\cal K}_{n-1}^- \otimes {\cal K}_n\subset {\cal F} \otimes {\cal K}_1 \otimes {\cal K}_2 \ldots \otimes {\cal K}_n = {\cal H}.
\end{eqnarray}

The   sequence of projectors
$\hat{Q}_i$ associated to ${\cal H}_i$ is
\begin{eqnarray}
 \hat{1} \otimes (1 - \hat{E}_1) \otimes (\hat{1} - \hat{E}_2) \otimes \ldots (\hat{1} - \hat{E}_n) < \hat{1} \otimes \hat{1} \otimes (\hat{1} - \hat{E}_2) \otimes \ldots (\hat{1} - \hat{E}_n)
 \nonumber \\
 < \hat{1} \otimes \hat{1} \otimes \hat{1} \otimes \ldots (\hat{1} - \hat{E}_n) < \ldots < \hat{1} \otimes \hat{1} \otimes \hat{1} \otimes \ldots \otimes \hat{1}.
\end{eqnarray}

The subspaces ${\cal V}_i$, defined by Eq. (\ref{vi}), are

 \begin{eqnarray}
 {\cal V}_i = {\cal F} \otimes {\cal K}_1 \otimes \ldots \otimes {\cal K}_i^+ \otimes \ldots \otimes {\cal K}_n,
 \end{eqnarray}
 with associated  projectors
\begin{eqnarray}
 \hat{P}_i = \hat{1} \otimes \hat{1} \otimes \ldots \otimes \hat{E}_i \otimes \ldots \otimes \hat{1}.
 \end{eqnarray}

\subsection{The macroscopic description of the detector}

A model for a measuring apparatus requires both a quantum and a classical description. The quantum description is necessary for  modeling the interaction between  the apparatus and the  measured system. The classical description is necessary for obtaining definite  measurement outcomes. At the macroscopic level, the most important parameter is the detector's macroscopic motion in Minkowski spacetime, which is described classically.

We assume that the spatial extension of the detector at rest corresponds to a subset $S$ of Euclidean space ${\bf R}^3$. We represent the points of $S$ by three-vectors ${\bf q}$. We denote the maximal dimension of the detector in its rest frame by $L$, i.e.,
\begin{eqnarray}
L = \sup_{{\bf q},{\bf q'} \in S}|{\bf q} - {\bf q'}|. \label{lqq}
\end{eqnarray}
A moving detector corresponds to a world-tube
 $S \times {\bf R}$ in Minkowski spacetime,  described by a spacelike embedding function

\begin{eqnarray}
{\cal E}: S \times {\bf R} \rightarrow M. \label{embed}
\end{eqnarray}

 The embedding is expressed in terms of coordinate functions ${\cal E}^{\mu}(\tau, {\bf q})$, for $\tau \in {\bf R}$ and ${\bf q} \in S$. Given the embedding function, we define the frame vector fields on the world-tube
 \begin{eqnarray}
 \dot{\cal E}^{\mu} = \frac{\partial {\cal E}^{\nu}}{\partial \tau}  \hspace{1cm} {\cal E}^{\mu}_i = \frac{\partial {\cal E}^{\nu}}{\partial q^i}.
 \end{eqnarray}

 The time-variable $\tau$ can be chosen as the proper time
  of the path ${\cal E}^{\mu}(\tau, q)$ for fixed ${\bf q}$, i.e., so that
\begin{eqnarray}
\eta_{\mu \nu}  \dot{\cal E}^{\mu}  \dot{\cal E}^{\nu} = -1.
\end{eqnarray}

We note that for a detector moving along inertial worldlines, the embedding functions are linear functions of ${\bf q}$ and $\tau$,
\begin{eqnarray}
{\cal E}^{\mu} = \dot{\cal E}^{\mu}\tau + {\cal E}^{\mu}_i q^i, \label{lorentz}
\end{eqnarray}
and the frame fields $\dot{\cal E}^{\mu}$ and ${\cal E}^{\mu}_i$ define a Lorentz transformation.

\subsection{The choice of the time parameters}
When constructing the probabilities associated to a sequence of $n$ measurements, a   crucial issue  is to select  the time variables $t_i$, with respect to which the probabilities (\ref{npp}) are defined. There are two constraints in this choice. First,  all time variables must eventually make reference to some time coordinate $t$ of Minkowski spacetime. This is because  the probabilities explicitly depend on the time-ordering of events and the time ordering is determined by the  causal structure of the spacetime. Second, in Poincar\'e-covariant QFTs, inertial reference frames   are distinguished.

One possibility is to use Eq. (\ref{npp}) with all events labeled by the same  coordinate time as measured in a Lorentz  frame. However, the resulting probabilities are explicitly dependent on the arbitrary choice of the time coordinate, and any discussion of Lorentz covariance becomes highly complicated.

   The covariance properties of the probabilities would be much simpler if we could employ a single
  proper time coordinate $\tau_i$ for each detector, so that  an one-to-one relation between each $\tau_i$ and a Lorentz coordinate $t$ exists.
      This is impossible unless the detector is strictly pointlike. To see this, consider an embedding, such that ${\cal E}^0$ is ${\bf q}$ independent in one  coordinate system. In this system, there exists  an one-to-one  function between the proper time variable $\tau$ and the coordinate time $t = {\cal E}^0(\tau)$. However, in any other coordinate system, obtained from the first through a Lorentz boost, the proper-time variable is ${\bf q}$-dependent. In general, there is a different proper time for each path  ${\cal E}^{\mu}(\cdot, {\bf q})$ within the detector's world tube.
   This is nothing  but the classic problem of relativistic simultaneity: there is no preferred time parameter by which to define simultaneity in an extended system.

This problem is resolved for sufficiently small detectors, by taking
  the detector's temporal coarse-graining  into account. Let us select a point $O$ as the `center' of the detector, conveniently taken at ${\bf q} = 0$. We identify the proper time associated to the path of $O$ as the proper time of the detector.  In the rest frame of the detector, the ambiguity in the definition of a single proper time parameter is of the order of $L$, Eq. (\ref{lqq}). Let us denote by $\sigma$ the coarse-graining time-scale in the rest frame. Then, the ambiguity in the definition of proper time due to relativistic non-simultaneity is negligible,  if
\begin{eqnarray}
L << \sigma. \label{nons1}
\end{eqnarray}

Next, we consider the same detector in a state of motion. Its world-tube is described by an embedding function ${\cal E}$. Two events separated in proper time by $\delta \tau$, are separated by the timelike vector $\dot{\cal E}^{\mu} \delta \tau$.
Two points of the detector separated by $\delta {\bf q}$ in the rest frame are separated by a spacelike vector ${\cal E}^{\mu}_i \delta q^i$. We assume that $\delta \tau = \sigma$, and choose $\delta {\bf q}$ such that $|\delta {\bf q}| = L$. Then,  the non-simultaneity of points in the detector can be ignored
\begin{eqnarray}
\eta_{\mu \nu} \dot{\cal E}^{\mu} \dot{\cal E}^{\nu} \sigma^2 >>|| h|| L^2, \label{nons2}
\end{eqnarray}
where $||h||$ is the norm of the $3 \times 3$ positive matrix $h_{ij} = {\cal E}^{\mu}_i {\cal E}^{\nu}_j  \eta_{\mu \nu}$.

 Eqs. (\ref{nons1}) and (\ref{nons2}) are sufficient for assigning a unique proper time $\tau$ to the detector that is in one-to-one correspondence with any Lorentzian time coordinate $t$. Thus, in any Lorentzian reference frame, the zero-th component ${\cal E}^0(\tau, {\bf q})$ can be approximated by ${\cal E}^0(\tau) := {\cal E}^0(\tau, 0)$. We will denote the inverse function that expresses $\tau$ in terms of $t$ as $\tau(t)$.

\subsection{Dynamics}

The Hamiltonian of the total system that includes the microscopic particles and the detectors consists of a free part and an interacting part. In the present context, the word "free" means that this part of the Hamiltonian involves no interaction between the quantum field and the detectors, not that the QFT under consideration describes free particles. The method fully applies to interacting QFTs.

We choose an arbitrary Lorentz frame with co-ordinate $t$, and we denote by $\hat{H}_{\phi}$ the associated Hamiltonian operator on the Hilbert space
   ${\cal F}$.
We assume that each detector is described by a Hamiltonian operator $\hat{h}_i$, defined on the Hilbert space ${\cal K}_i$. The operator $\hat{h}_i$ generates translation with respect to the proper time $\tau_i$ of the $i$-th detector. The evolution operator leaves the subspaces ${\cal K}_i^{\pm}$ invariant, i.e., it does not generate  transitions from ${\cal K}_i^-$ to  ${\cal K}_i^+$.

Hence, the free part of the Hamiltonian corresponds to the evolution operator on the Hilbert space ${\cal H}$ of the total system
\begin{eqnarray}
\hat{U}(t) = e^{-i\hat{H}_{\phi}t} \otimes e^{-i\hat{h}_1 \tau_1(t)} \otimes \ldots \otimes e^{-i\hat{h}_n \tau_n(t)}.
\end{eqnarray}

The interaction term $\hat{H}_I$  is  responsible for the transitions
associated to  measurement events
$\hat{H}_I$ is a sum of $n$ operators $\hat{V}_i$, each corresponding to a separate interaction of a the $i$-th detector with the quantum field,
\begin{eqnarray}
\hat{H}_I = \sum_{i=1}^n \hat{V}_i. \label{hint11}
\end{eqnarray}

There is no interaction between the detectors. Causality    implies that the operators $\hat{V}_i$ are local functionals of the field operators,

\begin{eqnarray}
 \hat{V}_i = \int d^3 x \hat{Y}_A({\bf x}) \otimes \hat{1} \otimes \ldots \otimes \hat{J}_i^A({\bf x}) \otimes \ldots \otimes \hat{1}.
  \label{vterm}
\end{eqnarray}
 where $\hat{Y}_A({\pmb  x})$ is a composite operator on ${\cal F}$ that is a local functional of the fields $\hat{\phi}_r$, and $\hat{J}^A_i({\bf x})$ are current operators  defined on the Hilbert space ${\cal K}_i$ of the $i$-th detector. $A$ is a collective index for the composite operators.

 The specific form of the composite operator depends on the physical processes involved in the detection. Consider, for simplicity, the case of a single free field $\hat{\phi}_r$---$\hat{\phi}_r$ may be scalar, spinor or vector. If $\hat{Y}_A \sim \phi_r$, then the interaction between field and detector is linear with respect to the creation and annihilation operators of the field. The action of the creation operator is suppressed in measurement, so the detection proceeds by annihilation (absorption) of a particle in the detector. If $\hat{Y}_A \sim \phi_r^2$, then the dominant contribution to the detection probability comes from terms with one creation and one annihilation operator, which correspond to particle scattering.

\subsection{Initial state of the detector}
We   assume a factorized initial state for the total system including the quantum field and the detectors, i.e., a state of the form
\begin{eqnarray}
|\psi\rangle_0 \otimes |\omega_1\rangle \otimes |\omega_2 \rangle \otimes \ldots \otimes |\omega_n\rangle \in {\cal H}. \label{factorizedpsi}
\end{eqnarray}
 A factorized initial state between detector and apparatus is commonly assumed in most models of quantum measurement theory. The measured system and the apparatus are assumed to be non-interacting prior to measurement, so there is no dynamical generation of correlations at any other stage other than the measurement.

In QFT, a generic state of the system does involve correlations between field and apparatus, because their interaction cannot be switched off. One expects that the initial state of the apparatus is "dressed" with vacuum fluctuations of the field, which  induce a renormalization of the physical parameters of the detector. However, in any reasonable modeling of a measurement apparatus, dressing should not affect the  correlation between pointer variables and   microscopic degrees of freedom. Its effect should be included into the noise that characterises the evolution of any coarse-grained observable \cite{GeHa}.

Hence, we expect that the consideration of factorized initial states is an approximation that does not significantly affect the probabilities associated to measurements. Renormalization will be needed, because, strictly speaking, the vector (\ref{factorizedpsi}) does not belong in the  Hilbert space where  a Hamiltonian with  the interaction term (\ref{hint11}) exists.  This is not   an issue in our models, because by employing the class operators  we work in the lowest order of perturbation theory---all probabilities are calculated at the tree level. The factorization approximation, Eq. (\ref{factorizedpsi}), might lead to small terms in the probabilities that violate causality, but these  correspond to higher order corrections that lie within the error margin of the  approximation.

In many models of quantum measurements, the initial state of the apparatus is  an eigenstate of the self-Hamiltonian of the apparatus, or close to such an eigenstate, so that it does not change prior to the interaction with the measured system.  This is a natural assumption for a static detector even in the relativistic set-up. However, a generalization is needed when dealing with moving detectors.

The necessary generalization is a  {\em stationarity condition} for   the detector. The stationarity condition involves the initial state $|\omega \rangle \in {\cal K}$, the self-Hamiltonian $\hat{h}$, the currents $\hat{J}^A({\bf x})$ of Eq. (\ref{vterm}) and the embedding ${\cal E}$ associated to a detector.
\begin{eqnarray}
\hat{J}^A({\bf x}) e^{-i \hat{h} \tau} |\omega\rangle = \int d^3q \hat{J}^A({\bf q}) \delta^3(x^i - {\cal E}^i ({\bf q}, \tau))|\omega'\rangle, \label{jcu}
\end{eqnarray}
for some vector $|\omega'\rangle$ and current operators $\hat{J}^A({\bf q}) $ defined on $S$.
The stationarity condition (\ref{jcu}) implies that
 the current operators and the initial state are combined in such a way such that the only time dependence  of $\hat{J}^A({\bf x}, \tau) e^{-i \hat{h} \tau}|\omega\rangle $ is due to the motion of the apparatus.

 For a static detector, Eq. (\ref{jcu})   means that $|\omega \rangle$ is an eigenstate of $\hat{h}$, and the corresponding energy eigenvalue is conveniently chosen to be zero.
    For a moving detector, Eq. (\ref{jcu}) means that the Hamiltonian  $\hat{h}$ affects only the part of the quantum state that corresponds to the apparatus'    macroscopic motion. In a point-like detector, Eq. (\ref{jcu}) reduces to a local field-particle  coupling that is commonly employed in Unruh-Dewitt detectors.

\subsection{Observables}

We represent the observables of the $i$-th detector by positive operators $\hat{F}_i(\lambda_i)$   defined on the subspace ${\cal K}_i^+$, such that

\begin{eqnarray}
\sum_{\lambda_i} \hat{F}_i(\lambda_i) = \hat{E}_i
\end{eqnarray}

The corresponding positive operators $\hat{\Pi}_i(\lambda_i)$ on the Hilbert space ${\cal H}$ of the total system  are
\begin{eqnarray}
 \hat{\Pi}_i(\lambda_i) = \hat{1} \otimes \hat{1} \otimes \ldots \otimes \hat{F}_i(\lambda_i) \otimes
\ldots \otimes \hat{1}. \label{Fpos}
 \end{eqnarray}

Since a detection event is localized in space, the set of measurement outcomes always includes the location of the detection event. Hence, the parameter $\lambda$ in the positive operators $\hat{F}(\lambda)$ is a shorthand for $({\bf Q}, \mu)$ where ${\bf Q}$ is a pointer variable that correlated to the coordinate ${\bf q}$ of the detector, and $\mu$ refers to pointer variables for magnitudes other than position. We have dropped the index $i$ labeling the detectors, as it is not needed.

Both  $\mu$ and ${\bf Q}$ are highly coarse-grained variables, since they corresponds to macroscopic records. Hence, $\hat{F}({\bf Q}, \mu)$ can be expressed as the product $\hat{F}_1({\bf Q}) \hat{F}_2(\mu)$, where $\hat{F}_1({\bf Q})$ and $\hat{F}_2(\mu)$ are POVMs for the pointer variables ${\bf Q}$ and $\mu$. This is because the commutator between two sufficiently coarse positive operators is   small---see, for example the coarse-grainings in Ref. \cite{Omn88}---  in the sense that
\begin{eqnarray}
\frac{Tr| [\hat{F}_1({\bf Q}), \hat{F}_2(\mu)]|}{Tr|\hat{F}_1({\bf Q})| Tr|\hat{F}_2(\mu)| } << 1.
\end{eqnarray}
It follows that
\begin{eqnarray}
\int d^3Q \hat{F}_1({\bf Q}) = \hat{E} \hspace{1cm} \sum_{\mu} \hat{F}_2(\mu) = \hat{E}.
\end{eqnarray}

\section{Probability assignment for relativistic measurements}

In Sec. 3, we introduced our model for relativistic quantum measurements. Essentially, we identified the relevant Hilbert spaces and the operators that enter into the expressions that were derived using the QTP method in Sec. 2. In this section, we compute the associated probabilities and examine their properties.

\subsection{General properties}
Before proceeding to the explicit evaluation of probabilities for $n$ measurements, we first describe some  properties that are immediately evident from their method of construction. These properties depend only on the broad principles employed in our method and not in technical details or approximation scheme.

\begin{enumerate}
\item We employ the proper time associated to each detector as the time variables that label the measurement events. Furthermore,  the spatial location ${\pmb Q}$ of a record of observation is among the measured variables. Hence, the probability densities of Eq. (\ref{npp}) are of the form
    \begin{eqnarray}
    P(\tau_1, {\pmb Q}_1, \mu_1;\tau_2, {\pmb Q}_2, \mu_2;\ldots; \tau_n, {\pmb Q}_n, \mu_n ) \label{probden4}
    \end{eqnarray}
    where $\mu_i$ refer to any other observable that may be recorded. Using the embedding functions, the probabilities may be expressed in terms of the spacetime coordinates $X^{\mu} =  {\cal E}^{\mu}(\tau, {\pmb Q})$ as    $P(X_1, \mu_1;  X_2, \mu_2; \ldots; X_n, \mu_n)$, but this is just a matter of convention. The probabilities depend explicitly and non-trivially on the embedding functions, because the latter incorporate all effects due to the motion of the detector.

   % We note that the incorporation of all information about the motion of the detector into the embedding function can straightforwardly be generalized for curved spacetimes.

  \item Eqs. (\ref{npp}),   (\ref{todn}) and (\ref{vterm}) imply that the probability density  $P(\tau_1, {\pmb Q}_1, \mu_1;\tau_2, {\pmb Q}_2, \mu_2;\ldots; \tau_n, {\pmb Q}_n, \mu_n )$ for $n$ measurement events is a linear functional  of the $2n$-point correlation function
      \begin{eqnarray}
 G(X_1, A_1; \ldots, X_n, A_n| X_1', B_1; \ldots; X_n', B_n) = \nonumber \\
 Tr
\left[T[\hat{Y}_{A_n}(X_n) \ldots \hat{Y}_{A_1}(X_1)] \hat{\rho}_0  \bar{T}[\hat{Y}_{B_1}(X') \ldots \hat{Y}_{B_n}(X'_n)] \right], \label{2np}
\end{eqnarray}
 where   $\hat{\rho}_0$ is the initial state of the quantum field,    $X = ({\pmb x}, t)$ and $\hat{Y}_A(X) = \hat{Y}_A({\pmb x}, t) = e^{i\hat{H}_{\Phi}t} \hat{Y}_A({\pmb x}) e^{-i\hat{H}_{\Phi}t}$ is the Heisenberg picture version of the composite operator that appears in the interaction Hamiltonian (\ref{vterm}).

 The correlation function Eq. (\ref{2np}) has $n$ time-ordered arguments ($T$) and $n$ arguments in reversed time order ($\bar{T}$).  Such correlation functions appear in the Schwinger-Keldysh or Closed-Time-Path (CTP) formalism of QFT  \cite{CTP, CaHu08}. The usual formulation of QFT in terms of a generating functional for time-ordered correlation functions is useful for treating scattering processes, as it is associated to the $S$ matrix that relates {\em asymptotic} in and out states. The CTP formalism is mostly employed in problems that require the calculation of probabilities or expectation values at finite times $t$, as in the present work. The correlation functions of the CTP formalism involve both time-ordered and anti-time-ordered products and they can be obtained from a generating functional through differentiation. The CTP generating functional is  a double Fourier transform of the decoherence functional of the decoherent histories approach  \cite{HPOAn}.

\item We use a collective index $a$ to stand for the pair $(X, A)$ in the composite operator $\hat{Y}_A(X)$. Then the correlation function (\ref{2np}) is expressed as $G_{a_1  a_2 \ldots a_n}{}^{a'_1, a'_2 \ldots a'_n}$, where lower indices correspond to time ordered operators and upper indices to anti-time-ordered. Then, the probability densities (\ref{probden4}) are of the general form
    \begin{eqnarray}
    P(1,2, \ldots, n) = G_{a_1  a_2 \ldots a_n}{}^{a'_1, a'_2 \ldots a'_n} R_{(1) a'_1}^{a_1}  R_{(2) a'_2}^{a_2} \ldots R_{(n) a'_n}^{a_n}, \label{shorth}
    \end{eqnarray}
    where  index contraction corresponds to an integral over $X$ and summation over $A$. Each matrix $R_{(i)}$ contains all information about the $i$-th detector (including its state of motion) and is a density with respect to the measured observables $\tau_i, {\pmb Q}_i$ and $\mu_i$.  The probability densities factorize with respect to the kernels $R_{(i)}$, because the apparatuses have been assumed non interacting and uncorrelated. Thus, the modeling of a complex measurement with $n$ apparatuses can be reduced to the construction of the kernels $R_{(i)}$ for each apparatus.

    \item We assume that the Heisenberg-picture composite operators  $\hat{Y}_A(X)$ transform covariantly under the action of the Poincar\'e group $\hat{U}(\Lambda, a)$, i.e.,
\begin{eqnarray}
\hat{U}^{\dagger}(\Lambda, a) \hat{Y}_A(X) \hat{U}(\Lambda, a) = C_A^B(\Lambda) \hat{Y}_B(\Lambda^{-1}X+a), \label{poincare2}
\end{eqnarray}
for some $C_A^B$ that are uniquely determined from the way the composite operator $\hat{Y}_A$ is expressed in terms of the basic fields $\hat{\phi}_a$.

Eq. (\ref{poincare2})   guarantees that the $2n$-point functions, Eq. (\ref{2np}) also transforms covariantly. Hence, the covariance of the probabilities depends only on the kernels $R_{(i)}$. If they are constructed in a way that respects Poincar\'e covariance, the probabilities will also be Poincar\'e covariant. We note that the simple covariance properties of the probabilities is a direct consequence of the choice of the proper times as time-ordering parameters. As such, it is conditional upon the validity  of Eqs. (\ref{nons1}) and (\ref{nons2}) that are entailed by this choice.

\end{enumerate}

\subsection{The detector kernel}
We evaluate the probability density  associated to $n$ detection events explicitly.
\begin{eqnarray}
P(\tau_1, {\pmb Q}_1, \mu_1;\tau_2, {\pmb Q}_2, \mu_2;\ldots; \tau_n, {\pmb Q}_n, \mu_n ) =  \nonumber \\  \sum_{A_1, \ldots, A_n} \sum_{B_1, \ldots, B_n} \int d^4X_1 \ldots dX_n \int d^4X'_1 \ldots dX'_n
G(X_1, A_1; \ldots, X_n, A_n| X_1', B_1; \ldots; X_n', B_n)\nonumber \\
\times R_{(1)}(X_1,A_1; X_1', B_1| \tau_1, {\pmb Q}_1, \mu_1) \ldots R_{(n)}(X_n,A_1; X_n', B_1| \tau_n, {\pmb Q}_n, \mu_n).  \label{longh}
\end{eqnarray}

Eq. (\ref{shorth}) is indeed a shorthand for Eq. (\ref{longh}). Each matrix $R_{(i)}$ of Eq. (\ref{shorth}) corresponds to a different {\em detector kernel}  $R(X,A; X', B| \tau, {\pmb Q}, \mu) $ that is  defined as
\begin{eqnarray}
R(t, {\pmb x}, A; t', {\pmb x'}, B| \tau, {\pmb Q}, \mu) =  \int ds   g_{\sigma}(s)\delta [t - {\cal E}^0(\tau +\frac{s}{2})] \delta [t' - {\cal E}^0(\tau -\frac{s}{2})] \nonumber \\
 \langle \omega | e^{i\hat{h}(\tau-\frac{s}{2})}\hat{J}^{B}({\bf x'}) \sqrt{\hat{F}}({\pmb Q}, \mu) e^{i\hat{h}s} \sqrt{\hat{F}}({\pmb Q}, \mu)  \hat{J}^{A}({\bf x})e^{-i\hat{h}(\tau+\frac{s}{2})}|\omega\rangle. \label{kernel1}
\end{eqnarray}
In Eq. (\ref{kernel1}), we wrote $X = (t, {\pmb x})$ and $X' = (t', {\pmb x}')$.

The detector kernel simplifies when the stationarity condition (\ref{jcu}) for the initial state is imposed.

\begin{eqnarray}
R(X, A; X', B| \tau, {\pmb Q}, \mu) &=& \int ds g_{\sigma}(s) \int_S d^3q \int_S d^3q'  \delta^4[ X - {\cal E}(\tau +\frac{s}{2}, {\pmb q})]  \delta^4[ X' - {\cal E}(\tau -\frac{s}{2}, {\pmb q}')] \nonumber
\\
&\times& \langle \omega'| \hat{J}^{B}({\pmb q}') \sqrt{\hat{F}}({\pmb Q}, \mu)
e^{i\hat{h}s} \sqrt{\hat{F}}({\pmb Q}, \mu)  \hat{J}^{A}({\pmb q})|\omega'\rangle. \label{kernel2}
\end{eqnarray}

  There is an asymmetry between the spatial and temporal coordinates of the detector in Eq. (\ref{kernel2}). It is due to  the asymmetric implementation  of approximations employed in Eq. (\ref{kernel2}). The observable ${\pmb Q}$ is correlated to the position coordinates ${\pmb q}$ on the world-tube. Assuming that the measurement of ${\pmb Q}$ is localized with a width $\delta$, for scales of observation much larger than $\delta$, we can substitute ${\pmb q} = {\pmb Q} + \frac{\pmb r}{2}$,  ${\pmb q}' = {\pmb Q} - \frac{\pmb r}{2}$ and substitute the double integral over ${\pmb q}, {\pmb q}'$ with a single integral over ${\pmb r}$. Thus, Eq. (\ref{kernel2}) becomes
\begin{eqnarray}
R(X, A; X', B| \tau, {\pmb Q}, \mu) &=& \int ds g_{\sigma}(s) \int_S d^3r   w_{\delta}({\pmb r})  \delta^4[ X - {\cal E}(\tau +\frac{s}{2}, {\pmb Q} + \frac{\pmb r}{2} )]  \delta^4[ X' - {\cal E}(\tau -\frac{s}{2}, {\pmb Q} - \frac{\pmb r}{2})] \nonumber
\\
&\times&\langle \omega'| \hat{J}^{B}({\pmb Q} - \frac{\pmb r}{2}) \sqrt{\hat{F}}_2(\mu)
e^{i\hat{h}s} \sqrt{\hat{F}}_2( \mu)  \hat{J}^{A}({\pmb Q}+ \frac{\pmb r}{2})|\omega'\rangle, \label{kernel3}
\end{eqnarray}
where $w_{\delta}({\pmb r})$ is a function analogous to $g_{\sigma}$, but defined with respect the spatial coordinates. For detailed derivation of Eq. (\ref{kernel3}) and further simplifications, see Ref. \cite{AnSav15b}.

\section{Discussion}

Eq. (\ref{longh}) together with the expressions (\ref{kernel2}) and (\ref{kernel3}) are the main results of this paper. They define  a probability density associated to $n$ measurements in a relativistic system, in which the spacetime coordinates of events are genuine random variables. Our results constitute a broad generalization of existing measurement models, such as the Glauber and the Unruh-Dewitt detector. In particular, our method  applies (i)  to any QFT and for any field-detector coupling, (ii) to the measurement of any observable, and (iii) to arbitrary size, shape and motion of the detector, as encoded in its associated embedding function ${\cal E}$.

The probability density for $n$ measurement events consists of two components. All information about the quantum fields is contained in a correlation function of a composite operator with $n$ time-ordered and $n$ anti-time-ordered entries. The information about each apparatus, including the measured variables and its state of motion,  is contained in the detector kernel, Eq. (\ref{kernel2}). The detector kernel is constructed unambiguously once the basic properties of the apparatus  have been specified.
In Ref. \cite{AnSav15b}, we study the detector kernel in some detail, and identify expressions that correspond to ideal measurements, i.e., forms of the detector kernel   that do not depend on the modeling details. In particular, we show how Glauber's photodetection theory, detectors of Unruh-Dewitt type, relativistic spin measurements and the QTP description of arrival time arise as particular cases of the formalism.

The presence of the delta functions in Eq. (\ref{kernel3}) implies that the $2n$-point functions (\ref{2np}) are evaluated with  arguments that correspond to the detector embeddings ${\cal E}$. Thus, the motion of the apparatuses is implemented in covariant way that can also be generalized for describing quantum fields curved spacetime. As long as the definition of the observables $\mu$ does not explicitly depend on the choice of a global coordinate system, the probability densities (\ref{longh}) are Poincar\'e invariant.

The  simple form of our result is due to the fact that we worked to lowest order in perturbation theory with respect to the field-detector coupling. For most applications this is sufficient. The leading  contribution contains the most important  physical characteristic of the measurement, namely, the correlation between microscopic variable and macroscopic record. Higher order corrections can be viewed as noise that obscures this correlation.

 Furthermore, the perturbative evaluation also allows us to sidestep complications due to the two problems that we encountered in the course of our derivation. These problems are (i) regularizing  the restricted propagator in order to guarantee positive definite probabilities (Sec. 2.3.2) and (ii) defining  `dressed' states for the apparatuses (Sec. 3.5). While these issues do not affect the predictions at our level of approximation, they must be successfully addressed in any theory  of  relativistic quantum measurements that makes a claim of conceptual completeness.

We believe that the formalism presented here provides a powerful working tool for bringing new insight into long-standing issues in the foundations of relativistic quantum physics, such as the minimum localizability of relativistic particles \cite{localization}, causality in the  two-atom system \cite{causality},  the identification of the correct operators for relativistic spin measurements \cite{spinop}, and understanding    `state reduction' and transmission of information in multi-partite  systems \cite{relred}.

 %This is an unphysical behavior, which originates from the unphysical assumption of the same absorption rate at all energies. A more realistic   assumption is to consider a constant detection rate up to a high-energy cut-off $\Lambda$.  Choosing an exponential cut-off $\alpha(\omega) \sim e^{-\omega/\Lambda}$.
 %Then we obtain
 %\begin{eqnarray}
 %\eta(s) = C' \mbox{Re}\left[ K_0(\Lambda^{-1} + i m s) + K_0(\Lambda^{-1} - i m s) \right]
 %\end{eqnarray}
 %in terms of the Bessel K-function  a positive constant $C'$.

\end{document}